\documentclass[12pt]{spieman}  
\usepackage{amsmath,amsfonts,amssymb}
\usepackage{graphicx}
\usepackage{setspace}
\usepackage{tocloft}
\usepackage{pdflscape}
\usepackage{comment}
\usepackage{url}

\title{Rapid far-infrared spectral-timing of X-ray binaries with PRIMA}

\author[a,*]{Alexandra J. Tetarenko}
\author[b]{Poshak Gandhi}
\author[c]{Devraj Pawar}

\affil[a]{Department of Physics and Astronomy, University of Lethbridge, Lethbridge, Alberta, T1K 3M4, Canada}
\affil[b]{School of Physics and Astronomy, University of Southampton, Southampton SO17 1BJ, UK}
\affil[c]{Department of Physics, R. J. College, Ghatkopar (W), Mumbai-86.}

\cftpagenumbersoff{figure}
\cftpagenumbersoff{table} 
\begin{document} 
\maketitle

\begin{abstract}
The most powerful cosmic engines in our universe are fueled by compact objects. These objects accrete large amounts of material and eject matter in the form of jets. Recent groundbreaking discoveries of gravitational waves from merging compact objects and the direct imaging of the black hole shadows with the Event Horizon Telescope, represent major steps forward in our understanding of such systems. However, there exists a large population of stellar-mass compact objects in our own Galaxy, present in X-ray binaries, which provide better laboratories with which to study the processes of accretion and ejection. X-ray binaries produce highly variable emission on timescales ranging from milliseconds (for light-travel time in the region close to the compact object) to weeks (governing the mass-inflow process). Therefore, high-time resolution observations can be a powerful tool to study these systems. However, as X-ray binaries emit across the electromagnetic spectrum, a suite of facilities is needed to take full advantage of these techniques. The PRIMA Observatory (PRobe far-Infrared Mission for Astrophysics) will provide unique access to a wavelength range which has not been sampled in X-ray binaries, representing an exciting new possibility for characterizing rapid time-domain phenomena of X-ray binaries (and potentially other transient sources) in the far-infrared regime.

\end{abstract}


{\noindent \footnotesize\textbf{*}Corresponding author,  \linkable{alexandra.tetarenko@uleth.ca} }

\begin{spacing}{2}   

\section{Introduction}
\label{sect:intro}  
X-ray Binaries (XRBs) are Galactic binary systems with a stellar-mass compact object (black hole or neutron star) accreting matter from a companion star, where a portion of the accreted material can be ejected in the form of a relativistic jet \cite{rem06,bah23}. {XRBs are further classified as {\it low-mass} or {\it high-mass} based on the mass of the companion star.} These binary systems are excellent laboratories to study the extreme physical conditions governing accretion and ejection as they represent nearby, rapidly evolving, and scaled-down analogues of their supermassive counterparts in active galactic nuclei (AGN).
Additionally, massive XRBs are potential progenitors of gravitational wave sources, and unlike most gravitational wave merger events, XRBs allow for long-term electromagnetic studies to track their system properties, natal kicks, and long-term evolutionary changes.\cite{gandhi19, atri19, zhao23}.

XRBs can either be transient, evolving from periods of minimal activity into bright outbursts on timescales of days to months, or more persistent in nature, in which case they remain in a bright, outbursting state for extended periods of time \cite{tetarenkob2015,blackcat}. In both instances, the XRB will cycle between different accretion states, in which the emission properties of the accretion flow (geometry, mass accretion rate) and the jet (brightness, morphology, spectral, and temporal properties) can vary dramatically. Accretion states are defined by the property of X-ray hardness, which is a measure of where the energy in the X-ray photons originates; high energy (hard) or low energy (soft) bands. A {\it ``hard state"} is dominated by non-thermal, high energy emission, and a {\it ``soft state"} is dominated by thermal, low energy emission.

For the transient XRBs, at the beginning of an outburst the source is typically in a {\it hard state}, with a compact jet and a geometrically thick, optically thin accretion flow \cite{naryi95,mar05}. As the X-ray luminosity (and accretion rate) rise, the source transitions from the {\it hard} to {\it soft} states, and the jet morphology changes from a compact jet to discrete plasma ejections. Once the source reaches the {\it soft} state, characterized by a geometrically thin, optically-thick accretion disc, the jet emission is quenched \cite{fen06,fengal14}. Following this sequence of events, the X-ray luminosity of the XRB will begin to fade. At later times, the source then undergoes a reverse transition (from {\it soft} to {\it hard}), where the spectrum hardens once again and the compact jet is re-established. The source remains in the hard state as it fades into quiescence. For the more persistent systems, similar accretion state changes are observed to those seen in the transients, but they tend to occur on much more rapid timescales (hours to days) and at more regular intervals {\cite{vanderklis, md14}.}

 \begin{figure}
\begin{center}
\includegraphics[height=8.5cm]{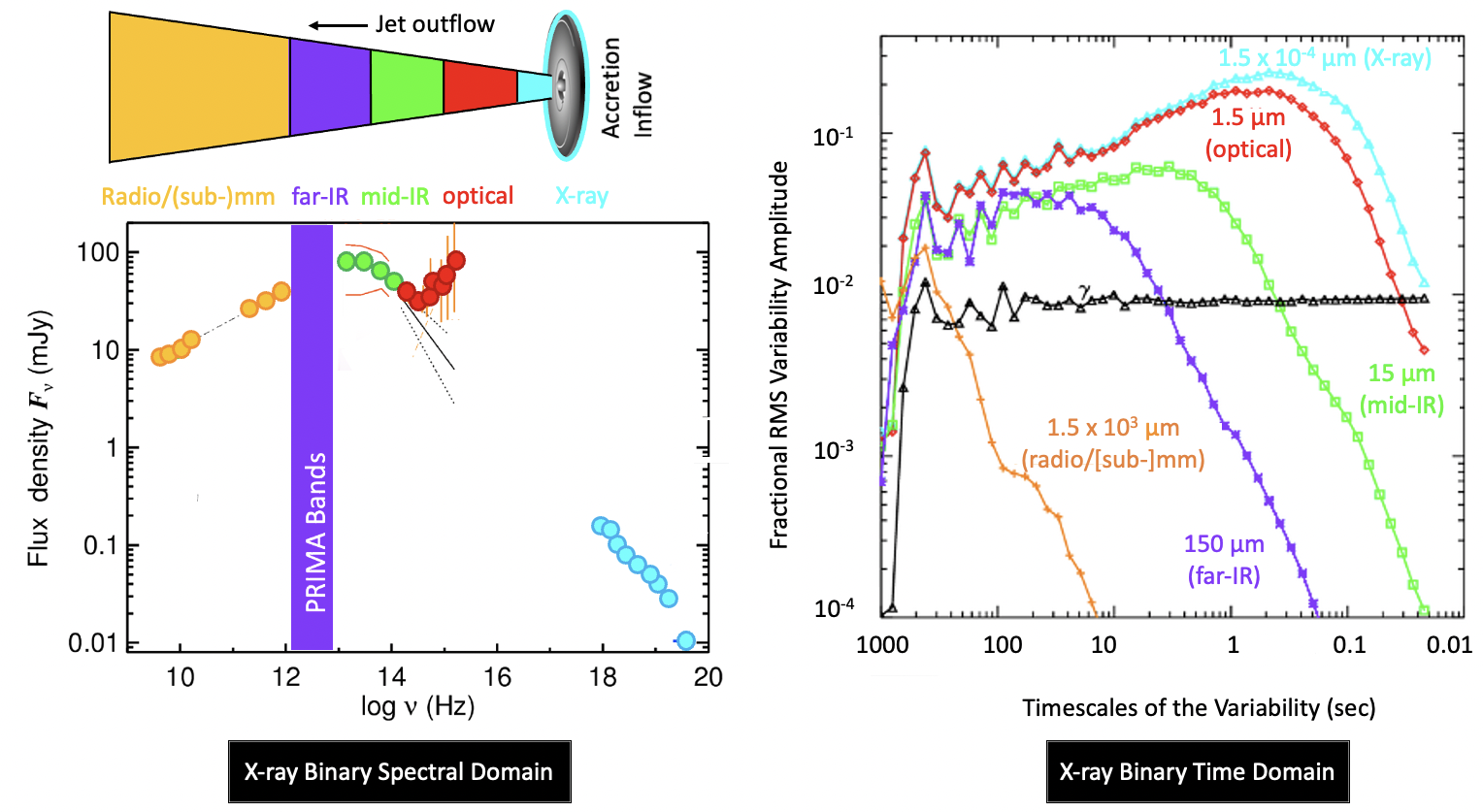}
\end{center}
\caption[Illustration of the XRB Spectral-timing Phase Space]
{ \label{fig:xrb} Illustration of the XRB spectral-timing phase space based on the canonical Galactic XRB source GX 339--4 \cite{gan11}. {\it Left:} A schematic of emission regions ({\it top}) and a snapshot of a characteristic XRB broad-band spectrum ({\it bottom}). {\it Right:} Fourier power spectra, characterizing the amplitude of emission variability from signals observed at five wavelength bands over different timescales ({calculated from a jet-based theoretical model \cite{mal14}, where the black curve represents the matter fluctuations injected into the jet}). In XRBs, shorter (longer) wavelengths probe closer to (further from) the compact object and show shorter timescale/larger amplitude (longer timescale/smaller amplitude) variations. The PRIMA bands fill in {an order of magnitude gap} in our broad-band wavelength coverage and are expected to show strong variability on timescales longer than about 0.1 seconds (see purple curve where the power spectrum breaks between $\sim0.1-10$ Hz). Linking simultaneous, multi-wavelength observations of XRBs, covering a range of timescales (sub-seconds to hours), with X-ray observations probing the in-flowing matter, is needed to measure how the variability signal propagates through the different emission regions.} 
\end{figure}

In XRBs, jet emission is produced as a result of synchrotron emission \cite{hj88aa,corfen02}, and displays a characteristic spectral shape characterized by a flat-to-inverted optically thick spectrum, ($\alpha\geq0, \textrm{where the emission intensity follows a }\hspace{0.1cm}F_{\nu}\propto\nu^{\alpha}$ relation), which transitions to an optically thin spectrum ($\alpha<0$) at a spectral break (Fig. 1 {\it bottom left}). This jet synchrotron spectrum is not static, but rather quite dynamic, evolving with accretion state changes (the spectral break shifts to longer wavelengths as the accretion rate increases \cite{r14,bag18,rus20,et24}). While jet emission dominates at longer wavelengths of light in XRBs (radio, sub-mm, infrared \cite{fen01,rus06,tetarenkoa2015,gandhip16}), emission from the accretion flow dominates at shorter wavelengths of light (UV, X-ray; Fig. 1 {\it top left} \cite{vanp1994,vanp96,d07}). The intermediate far-infrared/optical regime can contain contributions from both inflow and outflow, including jet synchrotron emission and reprocessed thermal emission from the outer accretion disc. These different emission processes are distinguishable, as they will operate on different timescales (jet emission will vary on faster timescales) and show different spectral shapes (reprocessed emission
should have a Rayleigh-Jeans $F_\nu\propto\nu^{+2}$  spectrum).

\section{Methodology of Spectral-timing Experiments for X-ray Binaries}
\label{sect:method}

Both the jet and accretion flow in XRBs produce highly variable emission in the spectral and temporal domains, and thus spectral-timing analyses offer new avenues to probe detailed jet and accretion properties (e.g., Refs.~\citenum{cas10,teta19b,vin23a}). In particular, spectral-timing can probe size-scales not accessible with current imaging capabilities. The smallest timescale over which the jet emission significantly varies (measured with Fourier power spectra) provides a direct measure of the jet size-scale at different wave-bands, enabling geometric measurements of the jet cross-section/opening angle (Fig. 1 {\it right}). As the jet propagates downstream from the compact object, optical depth effects cause longer wavelength emission to appear as a delayed version of shorter wavelength emission. Here the delay is directly related to jet speed and geometry. By modelling size-scale constraints and time-lag measurements between emission features at different wave-bands, constraints can be placed on the jet kinetic power and composition \cite{tet21}. Further, by cross-correlating multi-wavelength signals, accreting matter can be tracked from inflow to outflow, directly linking changes in the accretion flow with the jet at different scales.

Multi-wavelength variability measurements also allow for detailed tests of XRB models. In particular, one class of models predict that jet variability is driven by the injection of discrete shells of plasma injected at the base of the jet with variable speeds \cite{mal18}. The behaviour of these shells is directly linked to the amplitude of X-ray variability (traced by the X-ray power density spectrum; Fig. 1 {\it right} cyan curve). As the timescale of the jet variability depends on the shock speed and shell thickness, detecting correlated variability over a wide wavelength range could disentangle these parameters. Alternative models suggest variability could originate from cyclo-synchrotron emission produced in a precessing magnetized inner accretion flow \cite{vel13} or reprocessing of X-ray emission by the outer accretion disc \cite{vel15}. These models predict very different variability properties (e.g., anti-correlations between optical-infrared/X-ray,  and different spectral components present in the optical-infrared bands) when compared to the jet-based model, thus spectral-timing can be used to distinguish between the competing models, as well as investigate the interactions between the emitting components and their relative strengths.

While X-ray satellites were the first to pioneer spectral-timing studies of XRBs, new instrumentation has enabled the expansion of these studies into the optical, infrared, and radio wavelengths (e.g., Refs.~\citenum{ganh08,tet21,vinc18}). These studies have detected variability on timescales as short as $\sim0.1$ seconds in the optical-infrared, and tens of seconds in the radio regime, where these multi-wavelength variability signals show a high level of correlation with X-ray emission. Such experiments strongly suggest that variability in the accretion flow (probed through X-ray emission) is subsequently driving variability in the jet (probed through longer wavelength emission).

Followup studies have now successfully leveraged these new types of multi-wavelength time-series data sets to extract constraints on fundamental jet and accretion properties based on the variability signals alone. For example, Refs.~\citenum{tet21,zdz21} have measured jet power, speed, composition, and opening angle through detecting and modelling Fourier spectral properties and time-lags between time-series signals observed simultaneously at radio, optical-infrared, and X-ray wavebands in a black hole XRB source (Fig. 2). Optical-infrared studies have shown that Fourier domain techniques can probe physical processes in the jet and accretion geometry in the innermost regions near the compact object (e.g., Refs.~\citenum{cas10,vinc18,vinc21,ganh08,vin20,vin23b}). Additionally, radio and X-ray time-series observations have discovered a jet response to thermonuclear X-ray bursts occurring on the surfaces of accreting neutron stars in XRBs, and leveraged this discovery to measure jet speeds through a similar time-lag analysis \cite{rus24}. Overall, these results confirm the diagnostic potential of the spectral-timing experiments for studying jet and accretion physics in XRBs.

\begin{landscape}
 \begin{figure}
\begin{center}
\includegraphics[scale=0.35]{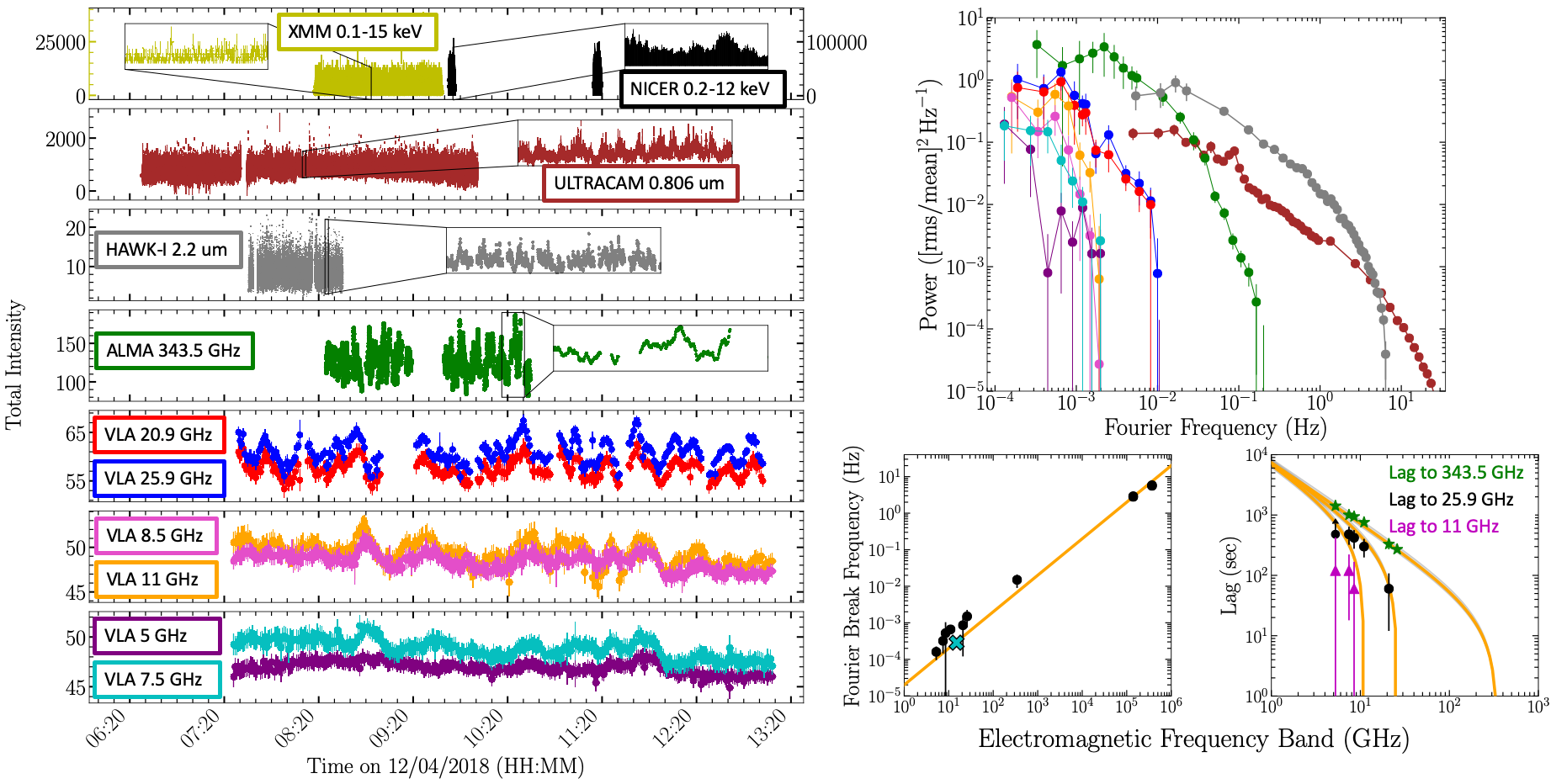}
\end{center}
\caption[Example Data Products Produced from Spectral-Timing Experiments]
{ \label{fig:spectimex} Example data products produced through a multi-wavelength spectral-timing experiment performed on XRB MAXI J1820+070 \cite{tet21,zdz21}. \textit{Left:} Multi-band light curves. \textit{Right top:} Fourier power spectra characterizing the variability amplitude of the light curves. An evolution in the shape of the power spectra is observed, where the break occurs at lower Fourier frequencies for signals at longer wavelengths (lower electromagnetic frequencies). This evolution likely traces the distance downstream from the black hole to the jet emission regions, allowing for the mapping out of the size-scales of the jet. \textit{Right bottom:} Modeling of jet timing properties \cite{blandford79}; Fourier power spectral breaks (the cyan marker represents a comparison to size-scales imaged at VLBI scales) and time-lags between bands (measured with cross-correlation analyses of light curves). Through simultaneously modeling these time-domain properties, the jet kinetic power, speed, and opening angle were measured.
{These results demonstrate a clear proof of concept, that fundamental XRB properties can be derived from time-series measurements alone.}} 
\end{figure} 
\end{landscape}

Despite the successful trajectory of spectral-timing experiments in the XRB field, there currently exist a number of limitations to these studies: (1) the far-infrared regime has never been sampled; (2) since the majority of studies are one-shot observations, they cannot probe how variability evolves as the spectrum/geometry of emission components change;
(3) since studies focusing on longer wavelengths have remained largely separate from those focusing on shorter wavelengths, connecting variability properties across different scales in the jet and accretion flow has not been possible; and (4) the state-of-the-art XRB variability emission models \cite{mal18,vel13,vel15,blandford79} have never been thoroughly compared and tested within the time-domain parameter space. 

The {\it PRIMAger hyperspectral imaging and polarimetry instrument} proposed for the PRIMA Observatory mission concept (see Glenn et al. and Ciesla et al., this volume) can overcome these limitations, as it is capable of providing sampling down to hundreds of milliseconds in twelve simultaneous filters across the far-infrared wavelengths. Further, this instrument should have sufficient sensitivity to observe tens of mJy-level objects, {typical of infrared emission levels in outburst accretion states \cite{bag18,echi24,gan25}}, and the ability for rapid response and flexible scheduling options which are optimal for targeting XRB sources. Lastly, as the far-infrared wavelengths represent the interface between jet-dominated and accretion flow-dominated emission bands, they span the waveband needed to disentangle the different variability models with PRIMA time-series measurements.

\section{Instrumental and Observational Requirements}
\label{sect:obsstrategy}

\subsection{Description of Observations}

The PRIMA Observatory mission concept is capable of performing a coherent census of far-infrared variability properties in XRBs through use of the {\it PRIMAger hyperspectral imaging and polarimetry instrument} to observe across different timescales and accretion regimes in a mixture of black hole and neutron stars. The {\it PRIMAger instrument} (see Ciesla et al., this volume) provides exquisite spectral coverage, as well as the ability to sample on rapid timescales. With these capabilities, PRIMA can track rapid spectral changes across the far-infrared band, enabling the cross-correlation of time-domain signals between the PRIMA filters (and with other multi-wavelength signals). Usage of the hyperspectral band is ideal here, as it maximizes the wavelength range sampled, covering 12 different filters from 25--80 microns. Observations of XRBs with the {\it PRIMAger instrument} would be most beneficial if undertaken with monitoring epochs of several hours on-source at a time resolution of 100 milli-seconds. With this time resolution, variability can be characterized at Fourier frequencies that exceed the expected break in the Fourier spectra by one order of magnitude (see Fig. 1 {\it right} purple curve), while staying well within the instrument’s capabilities. Observing for several hours on-source time per epoch ensures that multiple long-term variability cycles can be sampled and that PRIMA time-series can be compared with time-series signals from other facilities, to correlate far-infrared variability with variability at other wavelengths (weak correlated signals can usually be identified in a matter of hours; e.g., Ref.~\citenum{gan10}).

XRB targets are point-like, and thus PRIMA observations will not spatially resolve the system. Through utilizing PRIMA's small map mode for point source targets, the {\it PRIMAger instrument} is capable of sampling rates of 10 Hz (see Ciesla et al., this volume). Although, we note that this science can still be completed even if a compromise on filters vs. speed needs to be implemented based on further technical constraints of the instrument. Additionally, a sub-array mode could potentially enable even higher frame rates, and in turn open up new science windows for the brightest XRB targets (e.g., quasi-periodic oscillations [QPOs]; Ref.~\citenum{kalam16}). The small map mode ($\sim5\times5$ arcmin field of view) also allows for the target XRB source and at least one comparison star (for differential photometry) to be placed in the field of view. This procedure allows for verification that the variability detected with the {\it PRIMAger instrument} is intrinsic to the target XRB, and not due to instrumental effects, as the additional point source object within the field of view acts as a check source. These types of observations do not require absolute flux calibration, as the science is only driven by characterizing the relative emission intensity differences.

\subsection{Special Capabilities}
XRBs show drastic changes in emission across all wavelengths when they are in an outburst as a result of an unstable accretion rate onto the compact object. For instance, during these outbursts (which are usually not predictable and show emission intensity changes up to $\sim3-5$ magnitudes in the infrared/optical bands; e.g., Ref.~\citenum{gan11}) a range of rapid time-domain phenomena occur, such as periodic and aperiodic variability, bursts on time-scales of seconds, correlated spectral variability related to accretion dynamics in relativistic gravity, and radiative reprocessing. To best sample this range of time-domain phenomena, the following special capabilities incorporated into the PRIMA Observatory mission concept will be valuable for the successes of far-infrared spectral-timing observations of XRBs.

{\it Target of Opportunity and Rapid Response:} Since XRB outbursts are mostly unpredictable, the instrument time allocation policy should consider target of opportunity (ToO) proposals, with response times on the order of 24--48 hrs.

{\it High Cadence Monitoring:} Once in outburst the XRB source can remain above quiescent flux levels of most multi-wavelength facilities for weeks to months, during which multiple pointings at a variable cadence are needed (ranging from the fastest monitoring timescales on the order of a day to slower weekly monitoring timescales).

{\it Windowed Timing Mode:} Due to the compact nature of XRBs, they show rapid variability (up to kHz in neutron star systems), thus high time resolution photometry using instrument specific bands is the most suitable observation mode. Instruments with imaging CCDs designed for relatively longer integration times achieve this by using a subset of the total pixels on the CCD to enable rapid readout times. For example, such a strategy is achieved through the windowed timing mode (WT) on NASA’s Neil Gehrels Swift Observatory \cite{hill04,hill05}. A WT mode with variable time resolution will be very helpful for pushing the limits of far-infrared timing detections to sampling rates of $\sim100$ Hz or higher.

{\it Timing Accuracy:} To successfully perform far-infrared spectral-timing experiments, timing signals must be accurately compared across the PRIMA filters and to multi-wavelength signals obtained from other facilities. Thus relative and absolute timing accuracy is useful. Timing calibration of the {\it PRIMAger instrument} can be performed on a system with known timing properties, as has been done with the Neil Gehrels Swift Observatory \cite{cus12}, and more recently, the James Webb Space Telescope \cite{shaw24}. {For the case of the PRIMA observatory, optimal timing calibration sources may be those that undergo eclipses, which are often achromatic (a deep eclipsing system identified in the optical wavelengths may also be a good far-infrared target) and even fainter targets are feasible timing calibrators as significant signals can be built up by observing multiple eclipses. Additional possibilities may be magnetars, which have recently shown detectable pulsations in the sub-mm wavelengths \cite{torne22}, or possibly solar system objects with shorter rotation periods producing thermal/reflected emission in the far-infrared.}

{\it Synergy and Coordination:} To maximize science output of PRIMA far-infrared spectral-timing observations of XRBs, coordinating simultaneous multi-wavelength coverage with other facilities is necessary. For instance, the ground and space-based facilities that are currently capable of rapid timing observations include: radio (VLA, ATCA, MeerKAT), (sub-)mm (JCMT, ALMA, SMA, NOEMA, GBT), mid-infrared (JWST, VLT HAWK-I), optical (Gemini `Alopeke/Zorro, NTT ULTRACAM, GTC HiperCAM), and X-ray (XMM-Newton, NICER, Swift, INTEGRAL). Multi-wavelength programs designed for XRB spectral-timing which could potentially be paired with PRIMA time-series observations may take the form of 6-month/yearly target of opportunity programs and/or large multi-year, multi-facility programs (e.g., PITCH-BLACK\footnote{\url{https://www.eaobservatory.org/jcmt/science/large-programs/pitch-black/}}, GOFAST-XRB\footnote{\url{https://www.gemini.edu/observing/phase-i-proposing-time/llp/approved-llps/gofast-xrb-gemini-optical-fast-timing-x-ray}}).

\subsection{Integration Time Estimates}

For spectral-timing experiments of XRBs, the required integration time is not primarily dependent on sensitivity, but rather the need to observe long enough to catch several cycles of variability (typically on the order of hours at other wavelengths). Further, the total integration time will define the target brightness threshold, and in turn the sub-set of the XRB population reachable with PRIMA. Refs.~\citenum{mac19,vdk89} provide a prescription to quantify the detection of variability features in Fourier space at a certain significance based on the overall sensitivity achieved in the full observation time. 
\begin{equation}
S_{\rm var}=\frac{1}{2} S_{\rm total}^2 F_{\rm rms}^2 \left({\lambda}{t^{-\frac{1}{2}}}\right)
\end{equation}

Here, $S_{\rm var}$ represents the significance of features in variability (Fourier) space, $S_{\rm total}$ represents the signal-to-noise achieved in the full observation, $F_{\rm rms}$ represents the variability amplitude, $\lambda$ represents the width of variability in Fourier space, and $t$ represents the total observations time.

Based on the above prescription, assuming representative XRB values (i.e., $F_{\rm rms}=30$\%, $\lambda=4$ Hz, $t=10$ hours matching the PRIMA fact sheet; see Glenn et al., this volume), to achieve a $S_{\rm var}=10\sigma$ significance in Fourier space, a $S_{\rm total}=290\sigma$ detection is required in the full observation. Given a PRIMA point source sensitivity (at $5\sigma$) of $\sim1$mJy, this requirement sets the trigger threshold for our target sources to be $>50$ mJy. Based on the known XRB population, $\sim 1-2$ XRB sources meeting this trigger criterion should be observable per year with PRIMA ($\sim 63-86$\% chance of triggering every year).

\subsection{Target Selection}

To select the sample of target XRB sources ideal for observation with PRIMA, a balance can be struck between observational constraints (visibility and trigger threshold) and covering a wide range of the XRB parameter space (i.e., compact object mass and spin, accretion regime). This ensures PRIMA observations can be used to successfully detect and analyze far-infrared variability signals in the sampled XRB sources, as well as determine which XRB properties are the fundamental drivers behind far-infrared variability. As such, the following selection criteria could be used: (i) target sources must have known jet emission (through a clear radio detection), (ii) target sources must show levels of 
infrared emission detectable on short timescales with PRIMA (i.e., $\geq50$ mJy, XRBs typically reach these fluxes during outburst; see Fig. 1 {\it bottom left}), (iii) target sources must not be in complicated/crowded fields, and (iv) the target source pool must be a mix of black holes and neutron stars.

\subsection{Observing Campaigns}

As XRBs evolve over different timescales, showing a mix of accretion state change time-scales (transient XRBs spend on average 28 days in the hard state and 9 days transitioning to the soft state \cite{tetarenkob2015}), two types of observing campaigns would be useful to properly sample the far-infrared variability in XRBs with PRIMA. For all XRB sources targeted with PRIMA, X-ray and optical monitoring observations can be obtained (via X-ray satellites such as Swift, MAXI, NICER, and dedicated optical monitoring programs such as XB--NEWS \cite{xbnews}), to track accretion state changes in the targets.

{\it Short campaigns} can be used to target XRB sources showing evolutionary timescales on the order of hours--days. An example of a short campaign would consist of up to a week of observations with a daily cadence.

{\it Long campaigns} can be used to target XRB sources showing evolutionary timescales on the order of weeks--months. An example of a long campaign would consist of 10--20 observations with a cadence depending on the target, taken across multiple months. Here the target specific cadence may be defined by the target source evolution time-scales (as determined by radio, optical, and X-ray monitoring). This strategy ensures that PRIMA could sample different accretion states and the transition between states during outburst.

\section{Summary}

Stellar-mass compact objects in Galactic XRBs are ideal laboratories for the processes of accretion and ejection as they provide a real-time view of how the accretion flow and jets evolve and interact with their environment. As X-ray binaries are known to produce highly variable emission, time-domain analyses can be powerful tools to probe these complex physical processes.

X-ray satellites were the first to pioneer such timing studies of XRBs, but the recent invention of new instrumentation and observing techniques have now allowed for the expansion of these studies into the optical, infrared, and even the longest wavelength radio/sub-mm bands. The {\it PRIMAger instrument} in the PRIMA Observatory mission concept uniquely probes a wavelength range that has not been accessible so far (connecting sub-mm [probed by ALMA] and mid-infrared [probed by JWST]), with the capability to sample this emission on rapid timescales, and thus represents an exciting new possibility for characterizing XRB  variability.

To perform rapid time-domain science with PRIMA, two main operational capabilities are required; a rapid sub-second sampling mode and a triggered target of opportunity mode. With these capabilities, time-series analyses on PRIMA data will allow for accurate measurements of key physical properties in XRBs, and allow for detailed analysis of how matter propagates from inflow to outflow in these systems, through tracking how accretion flow variations propagate into the jet in different accretion states, and between systems housing different central compact objects (both black holes and neutron stars).  
Ultimately, through opening up time-domain science with PRIMA, observations of XRBs will act as a template for future transient observations and help to answer some of the most intriguing and impactful questions in the field today, including: (a) How are jets launched and accelerated? (b) What are the initial conditions for the formation of jets in the launching and acceleration region? (c) What role do mass, spin, nature of the accretor, accretion rate, and outburst duty cycle, play in jet production? (d) How do jet properties (energetics, jet speed, jet size scales) compare across systems housing different compact objects? (e) What factors drive jet evolution? (f) How does the energy released by jets compare to other feedback processes, such as star formation?

\section*{Disclosures}
The authors have no relevant financial interests in the manuscript and no other potential conflicts of interest to disclose.

\section* {Code, Data, and Materials Availability} 
There are no supporting data, codes, or materials for this manuscript.

\section* {Acknowledgments}
AJT acknowledges the support of the Natural Sciences and Engineering Research Council of Canada (NSERC; funding reference number RGPIN--2024--04458). DP acknowledges the support from ISRO-India through the RESPOND Programme. PG thanks the Royal Society Leverhulme Trust Senior Research Fellowship for support, together with the Science and Technology Facilities Council.


\bibliography{report}   
\bibliographystyle{spiejour}   


\section* {Biographies}

\vspace{2ex}\noindent\textbf{Alexandra J. Tetarenko} is an assistant professor at the University of Lethbridge. She received her BSc degree from the University of Calgary in Astrophysics, and her MSc and PhD degrees in Physics from the University of Alberta in 2014 and 2018, respectively. She is the author of more than 80 refereed journal articles, including 3 Nature and 2 Science papers. Her current research interests center on compact objects such as black holes and neutron stars.

\vspace{2ex}\noindent\textbf{Poshak Gandhi} is professor of astrophysics at the University of Southampton. He received his BSc Physics Honours from the University of Delhi, and his PhD in Astronomy from the University of Cambridge. He has published extensively on multiwavelength observational studies of accreting compact objects on all mass scales, and on astrometric studies of X-ray binaries. 

\vspace{2ex}\noindent\textbf{Devraj Pawar} is an associate professor in the Department of Physics at Ramniranjan Jhunjhunwala College in Mumbai. He has received his BSc degree from the University of Mumbai, his MSc from the University of Pune, and his PhD from the University of Mumbai. His research interests are multiwavelength studies of X-ray binaries.

\end{spacing}
\end{document}